\documentclass[journal]{IEEEtran}

\usepackage[numbers, sort&compress]{natbib}

\usepackage{amsmath, amssymb, amsfonts}
\usepackage{empheq}
\usepackage{booktabs}
\usepackage{threeparttable}
\usepackage{multirow}
\usepackage{multicol}

\usepackage[thinlines, thiklines]{easybmat}

\usepackage[linesnumbered, ruled, vlined]{algorithm2e}

\usepackage{graphicx, subfigure, svg}  

\usepackage[english]{babel}

\usepackage{arydshln}

\usepackage{enumerate}
\usepackage{paralist}
\usepackage{url}
\usepackage{color}

\usepackage{url}
\usepackage{hyperref}

\begin{document}

\renewcommand{\figurename}{Fig.}
\renewcommand{\tablename}{TABLE}

\title{Semi-Supervised RF Fingerprinting \\ with Consistency-Based Regularization}

\author{Weidong~Wang, Cheng~Luo, Jiancheng~An, Lu~Gan, Hongshu~Liao, and Chau Yuen, \IEEEmembership{Fellow, IEEE}}


\markboth{IEEE Internet of Things Journal}
{Wang \MakeLowercase{\textit{et al.}}: Semi-supervised RF Fingerprinting}

\maketitle

\begin{abstract}
    As a promising non-password authentication technology, radio frequency (RF) fingerprinting can greatly improve wireless security. Recent work has shown that RF fingerprinting based on deep learning can significantly outperform conventional approaches. The superiority, however, is mainly attributed to supervised learning using a large amount of labeled data, and it significantly degrades if only limited labeled data is available, making many existing algorithms lack practicability. Considering that it is often easier to obtain enough unlabeled data in practice with minimal resources, we leverage deep semi-supervised learning for RF fingerprinting, which largely relies on a composite data augmentation scheme designed for radio signals, combined with two popular techniques: consistency-based regularization and pseudo-labeling. Experimental results on both simulated and real-world datasets demonstrate that our proposed method for semi-supervised RF fingerprinting is far superior to other competing ones, and it can achieve remarkable performance almost close to that of fully supervised learning with a very limited number of examples.
\end{abstract}


\IEEEpeerreviewmaketitle

\section{Introduction} \label{Section: Introduction}
\IEEEPARstart{T}{he} internet of things (IoT) has gained a broader prospect when supported by modern wireless communication technology that can provide connectivity anytime and anywhere. Meanwhile, it shares more challenges from privacy security because wireless transmissions are open and accessible to both authorized and illegal users, leaving many IoT devices prone to malicious attacks like node forgery or impersonation \cite{zou2016survey}. The vulnerability of IoT devices in such network security issues has inspired research into physical layer security \cite{hamamreh2018classifications}, particularly radio frequency (RF) fingerprinting that aims to distinguish different transmitters by characterizing device-specific features (aka ``fingerprints") presented in their emitted signals \cite{xu2015device}, rather than considering those easily forgeable identifiers such as IP/MAC addresses or other user credentials that are often used in conventional authentication mechanisms.

Research on RF fingerprinting has practically been going on for decades. Conventional RF fingerprinting relies on careful engineering and considerable domain expertise to design suitable signal features, combined with classic machine learning techniques like support vector machines (SVM) for classification \cite{soltanieh2020review}. Recent breakthroughs in natural language processing, speech recognition, and computer vision have been made by deep learning \cite{lecun2015deep}, which also provides many new insights for RF fingerprinting. The powerful non-linear representation of deep neural networks makes it possible to learn potential high-level features directly from raw signal data with complex structures and inner correlations, which brings better performance and many other potential advantages. For example, RF fingerprinting based on deep learning also wins at scalability with wider model and protocol ranges and larger device populations \cite{jian2020deep}.

There have been many prior works for RF fingerprinting based on deep learning, but most were only limited to supervised learning, which requires massive training data, or said they often discussed all this with an ideal amount of labeled data provided. However, obtaining so much signal data with labels in many real-world IoT applications is challenging. The main reason is that an IoT system usually has a large number of devices. Such a process of signal acquisition and annotation itself is expensive and time-consuming. Training a deep neural network on small datasets generally leads to overfitting and thus significantly degrades its generalization ability. In particular, we may get worse results if some signal transmissions are mislabeled. The practicability of RF fingerprinting based on deep learning thus has been heavily restricted.

To this end, we develop a novel deep semi-supervised solution for RF fingerprinting, which employs data augmentation to increase a limited number of examples and simultaneously perturb massive unlabeled data for consistency-based regularization, resulting in a maximum generalization improvement. The main contributions are summarized as follows:
\begin{itemize}
    \item  We comprehensively discuss a variety of existing data augmentations for radio signals and analyze their effectiveness in RF fingerprinting. Moreover, we present a novel data augmentation scheme termed ``stochastic permutation", combined with signal rotation at specific angles to form composite data augmentation.
    \item  To fully exploit a considerable amount of unlabeled signal data that is easily available, we introduce a combination of consistency-based regularization and pseudo-labeling, effectively incorporated with our composite data augmentation to provide an end-to-end training pipeline for semi-supervised learning.
    \item  To achieve better performance, we carefully design a deep residual network (ResNet), empirically discuss how to design a deep neural network suitable for semi-supervised learning, and point out that a model should balance its complexity to compromise between required representation capacity and increased overfitting risk.
\end{itemize}

The proposed method for semi-supervised RF fingerprinting is verified using a series of experiments. The experimental results on both simulated and real-world datasets demonstrate that ours is far superior to other competing ones and only requires a small amount of labeled data to achieve almost fully supervised performance.

\vspace{11pt}

The rest of this paper is organized as follows. The background and related work are briefly described in Section~\ref{Section: Background and Related Work}. The analysis for data augmentation is provided in Section~\ref{Section: Data Augmentation}. The proposed method for semi-supervised RF fingerprinting is depicted in Section~\ref{Section: Methodology}. The experiments and related results are given in Section~\ref{Section: Experiments and Results} before concluding in Section~\ref{Section: Conclusion}.

\section{Background and Related Work} \label{Section: Background and Related Work}
Research in modulation recognition using deep learning \cite{o2016convolutional} has spurred renewed interest in many wireless communication issues. Refer to \cite{mao2018deep, sun2019application, pham2021intelligent} for more comprehensive reviews. Deep learning has also been introduced to RF fingerprinting with many remarkable efforts.

\subsection{Deep Learning for RF Fingerprinting}
Earlier, Riyaz \emph{et al.} \cite{riyaz2018deep} employed a simple convolutional neural network (CNN) for RF fingerprinting, taking raw signal waveforms (typically represented in an in-phase and quadrature form, i.e., I/Q) as input, which can perform better than conventional approaches. There are many similar studies, including those replacing I/Q with some image-like representations, e.g., differential constellation trace figures (DCTF) \cite{peng2019deep} and a variety of time-frequency distribution descriptions \cite{baldini2019assessment}. In particular, Yu \emph{et al.} \cite{yu2019robust} introduced a multisampling strategy to learn multiscale features.

In addition to various CNNs, many other network types have also been investigated for RF fingerprinting. The authors in \cite{roy2019rf} considered three kinds of recurrent neural networks (RNN) with different recurrent cells implemented, including long short-term memory (LSTM), gated recurrent unit (GRU), and convolutional LSTM. Different neural networks can meet various challenges. For example, since some rogue transmitters might produce fake signals to spoof trusted transmissions, Roy \emph{et al.} \cite{roy2019rfal} used a generative adversarial net (GAN) to learn an implicit sample space of known trusted transmitters, allowing its discriminator to distinguish real signals from those fake ones. In general, it has become a prevailing trend to exploit deep learning for RF fingerprinting.

\subsection{Deep Semi-Supervised Learning for RF Fingerprinting}
RF fingerprinting based on deep learning has made so many advances, but almost all belong to supervised learning and only consider an ideal amount of labeled data for training. It is usually difficult to achieve satisfactory results when considering a limited number of examples. The scarcity of labeled data is very common in practice since labeling data often requires human labor and costs a considerable amount of resources. The best practice to access this issue is to consider semi-supervised learning \cite{geffner2022introduction} that can leverage unlabeled data to improve generalization performance when only limited labeled data is available. There is rising interest in semi-supervised learning to combine deep neural networks, namely deep semi-supervised learning. The most remarkable impact has occurred in image classification, gradually extending to other fields \cite{chebli2018semi, chen2019semisupervised}.

As far as we are aware, deep semi-supervised learning has not yet been well investigated for RF fingerprinting, but there have been some efforts in another similar field of communication signal recognition, modulation recognition. Earlier, O'Shea \emph{et al.} \cite{o2017semi} employed convolutional autoencoders to learn embeddings from massive unlabeled signal data. The encoder part is then frozen and concatenated to a softmax classifier, fine-tuned with a small amount of labeled data. However, such embeddings, learned by minimizing reconstruction errors, are not necessarily applicable to classification. The two-stage training pipeline is also complex and easily accumulates more errors. There are also some studies based on data generation \cite{li2018generative, zhou2020generative}. The general idea is to exploit GANs for distribution learning, which can produce more examples to help improve generalization. Recent work by Dong \emph{et al.} \cite{dong2021ssrcnn} adopted a relatively modern design that simultaneously considers supervised and unsupervised losses.

In contrast to previous research, we argue that data augmentation can play an important role in semi-supervised RF fingerprinting. To this end, we design a composite data augmentation scheme for radio signals by analyzing their intrinsical characteristics, based on which we investigate semi-supervised RF fingerprinting in conjunction with consistency-based regularization and pseudo-labeling.

\section{Data Augmentation} \label{Section: Data Augmentation}
Domain knowledge is generally leveraged to design suitable data transformations leading to improved generalization, known as ``data augmentation". For example, one routinely uses rotation, translation, cropping, flipping, and random erasing to enforce visually plausible invariance in image classification \cite{shorten2019survey}. In this section, we shall introduce a variety of data augmentations used for radio signals, including our proposed ones, and analyze their effectiveness in RF fingerprinting.

\subsection{Analysis for Existing Data Augmentations}
Previous literature has proposed many data augmentations for radio signals, including rotation, flipping, noise adding, and GAN-based generation.

\vspace{10pt}

\noindent\textbf{Rotation and Flipping} Inspired by computer vision, Huang \emph{et al.} \cite{huang2019data} introduced rotation and flipping as data augmentation for radio signals. Given a radio signal $\boldsymbol{s}$, we can define its rotated version as
\begin{equation}
    \boldsymbol{s}_{\vartheta} = \operatorname{rot}_\vartheta(\boldsymbol{s}) = \boldsymbol{s} e^{j\vartheta}
\end{equation}
The angles for rotation ($\vartheta$) used in \cite{huang2019data} are chosen as $0^\circ$, $90^\circ$, $180^\circ$, and $270^\circ$ to improve performance for modulation recognition. The flipped version can be defined as
\begin{equation}
    \left\{
    \begin{aligned}
        \boldsymbol{s}_\mathrm{h} & = \operatorname{conj}(-\boldsymbol{s}) \quad \text{// Flip Horizontally} \\
        \boldsymbol{s}_\mathrm{v} & = \operatorname{conj}(\boldsymbol{s}) \quad \text{// Flip Vertically}    \\
    \end{aligned} \right.
\end{equation}
The function $\operatorname{conj}(\cdot)$ is to take conjugate. The essences of rotation and flipping, not mentioned in \cite{huang2019data}, are given below.

\begin{figure}[htb]
    \centering
    \subfigure[BPSK]
    {
        \centering
        \includegraphics[width=0.14\textwidth]{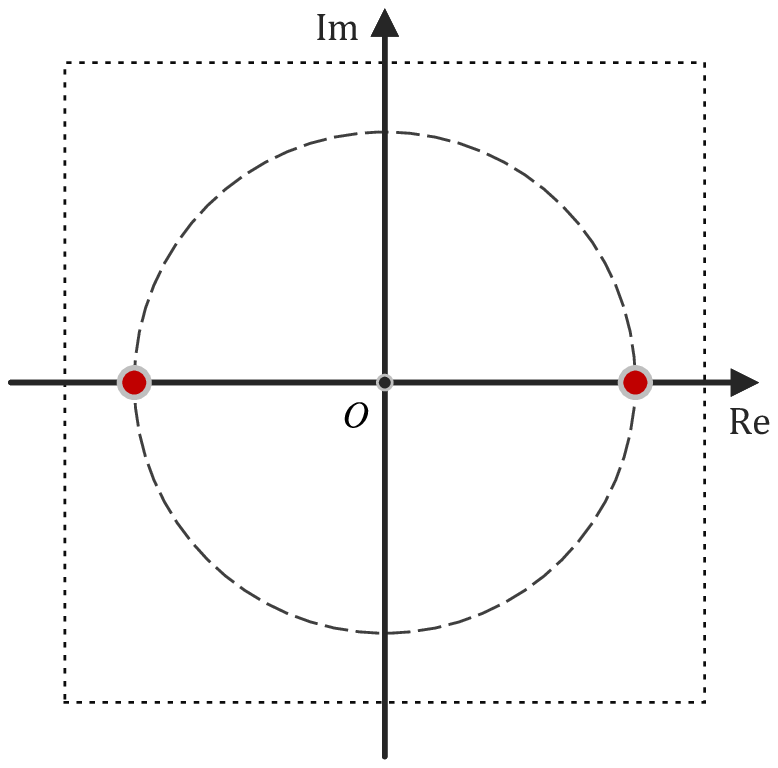}
    }
    \hfill
    \subfigure[QPSK]
    {
        \centering
        \includegraphics[width=0.14\textwidth]{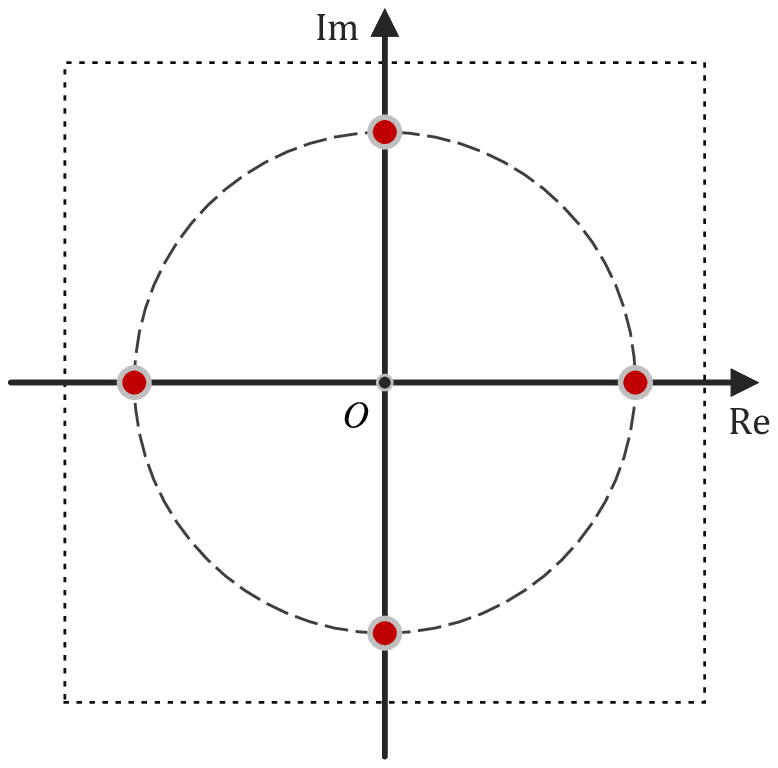}
    }
    \hfill
    \subfigure[8PSK]
    {
        \centering
        \includegraphics[width=0.14\textwidth]{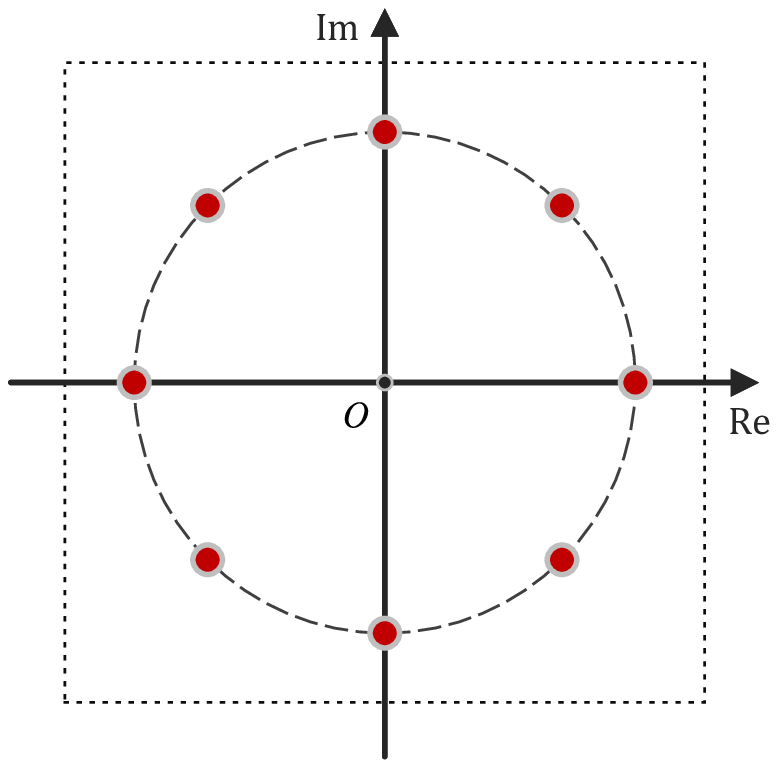}
    }
    \caption{Symbol topologies of BPSK, QPSK, 8PSK.}
    \label{Figure: Constellation Diagram}
\end{figure}

\begin{figure}[htb]
    \centering
    \subfigure[Rotate $0^\circ$ (Original)]
    {
        \centering
        \includegraphics[width=0.185\textwidth]{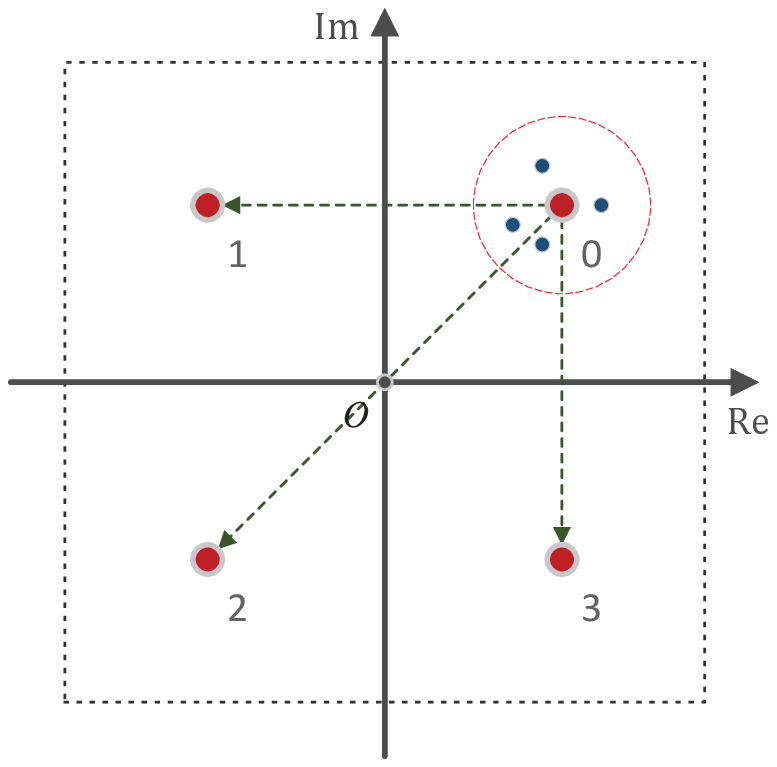}
    }
    \hspace{0.75em}
    \subfigure[Rotate $90^\circ$]
    {
        \centering
        \includegraphics[width=0.185\textwidth]{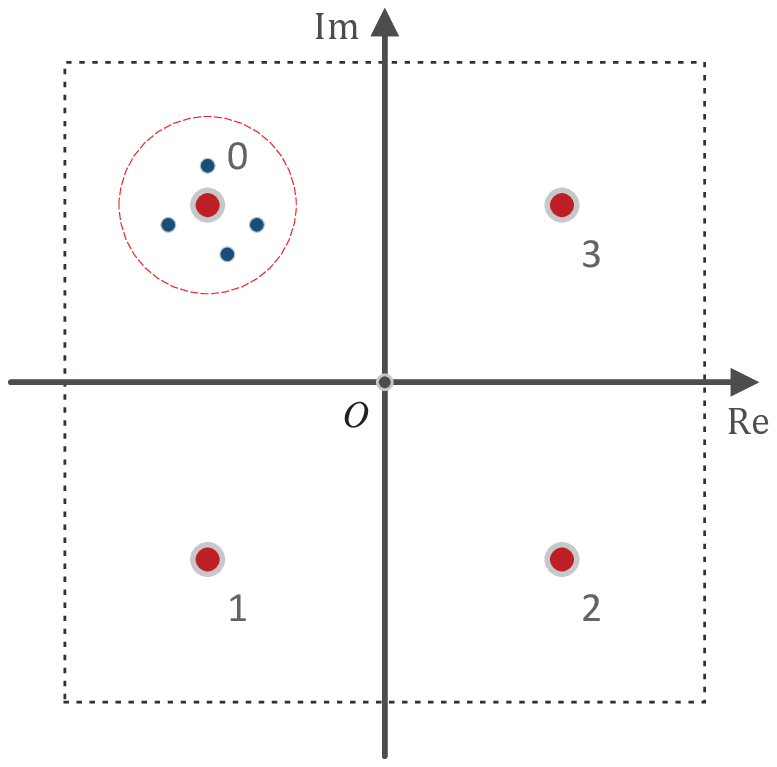}
    }
    \\
    \subfigure[Rotate $180^\circ$]
    {
        \centering
        \includegraphics[width=0.185\textwidth]{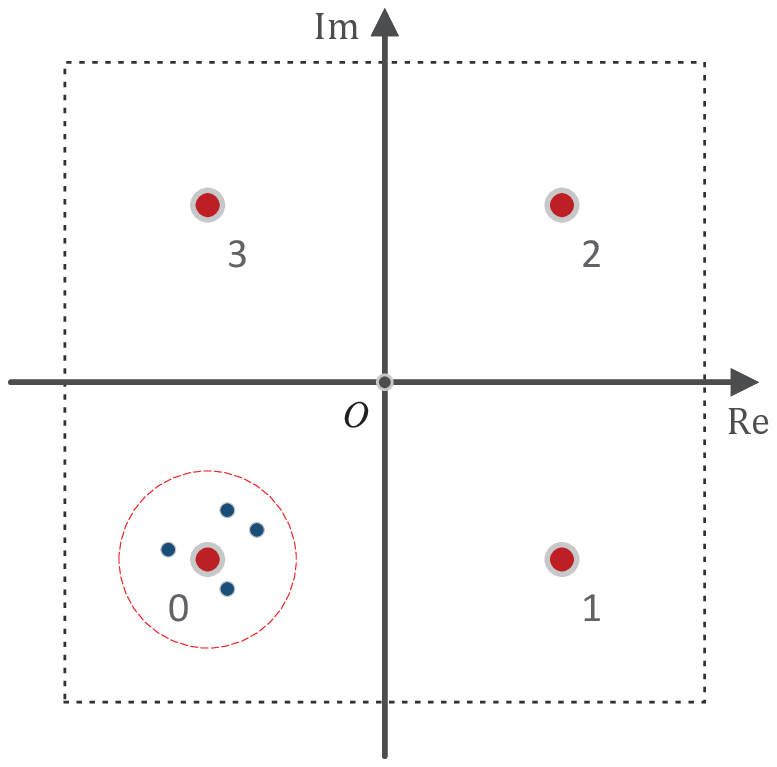}
    }
    \hspace{0.75em}
    \subfigure[Rotate $270^\circ$]
    {
        \centering
        \includegraphics[width=0.185\textwidth]{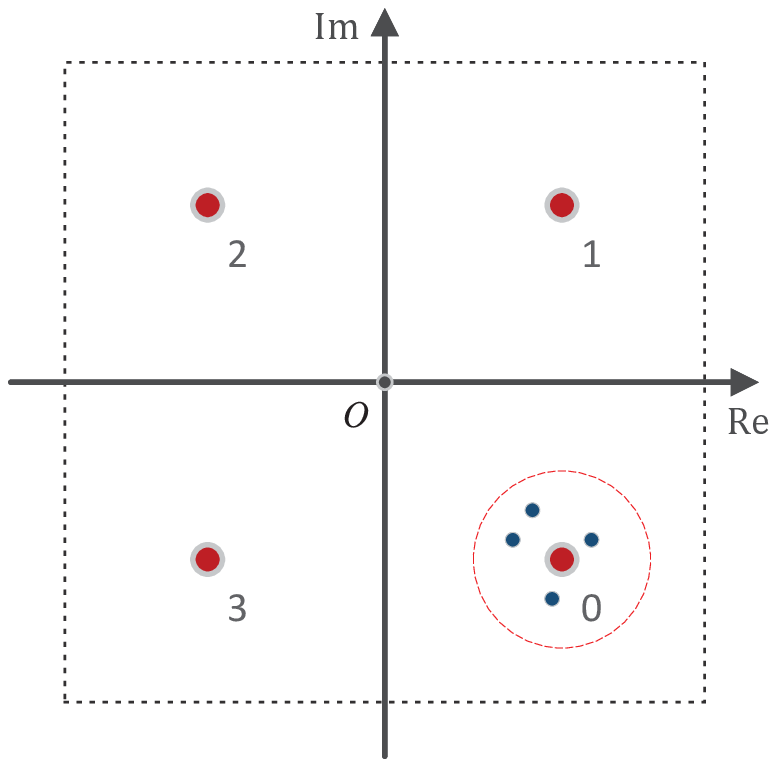}
    }
    \caption{Rotation of a QPSK-modulated signal.}
    \label{Figure: Rotation}
\end{figure}

As illustrated in Fig. \ref{Figure: Constellation Diagram}, we can see that most modulation schemes have a regular symbol topology. This characteristic determines that a radio signal can keep its corresponding constellation topology unchanged when rotated at some specific angles. For most cases (QPSK and M-QAM), such angles can be chosen as $0^\circ$, $90^\circ$, $180^\circ$, $270^\circ$, i.e., $\mathrm{rot}_{0^\circ}$, $\mathrm{rot}_{90^\circ}$, $\mathrm{rot}_{180^\circ}$, $\mathrm{rot}_{270^\circ}$, where $\mathrm{rot}_{0^\circ}$ actually obtains a radio signal itself. Take QPSK-modulated signals as an example. As illustrated in Fig. \ref{Figure: Rotation}, we can see that rotating at these specific angles allows one symbol and all of its belonging sampling points to be transformed into another, whereas any other angles cannot, only resulting in additional phase noise.

The noise components of different sampling points are often uncorrelated or partially correlated so that we can treat each of those rotated versions as an entirely new signal. Likewise, we can rotate BPSK-modulated signals only at $0^\circ$ and $180^\circ$, or rotate 8PSK-modulated signals at an interval of $45^\circ$ to have $8$ rotated versions. Since modulation schemes can often be considered a known parameter when considering RF fingerprinting, it may be possible to individually customize a set of rotation operations for each device according to their specific modulation schemes, which aims to achieve a maximum data augmentation effect.

\begin{figure}[htb]
    \centering
    \subfigure[Flip Horizontally]
    {
        \centering
        \includegraphics[width=0.185\textwidth]{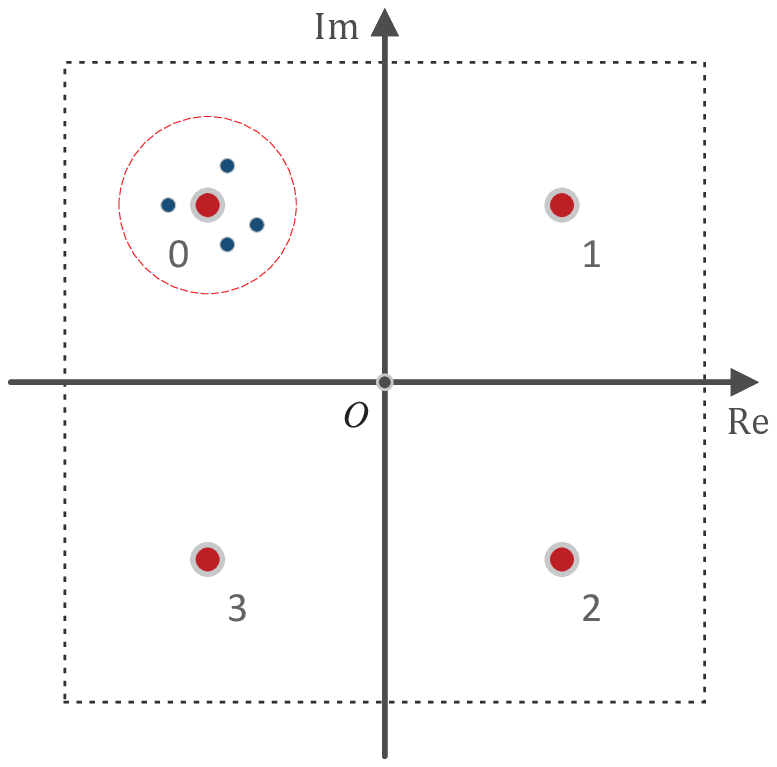}
    }
    \hspace{0.75em}
    \subfigure[Flip Vertically]
    {
        \centering
        \includegraphics[width=0.185\textwidth]{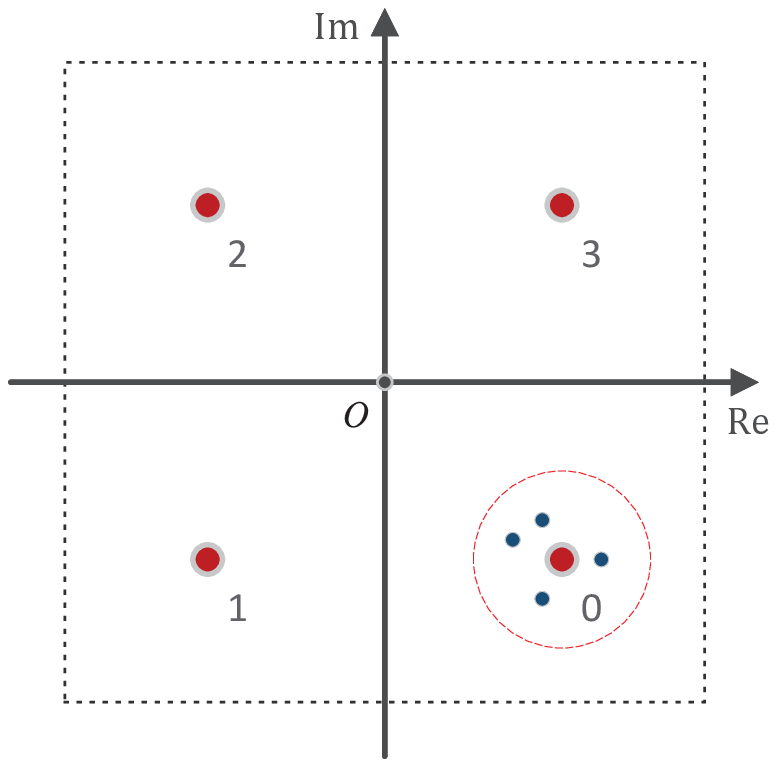}
    }
    \caption{Flipping of a QPSK-modulated signal.}
    \label{Figure: Flipping}
\end{figure}

\begin{figure}[htb]
    \centering
    \includegraphics[scale=0.685]{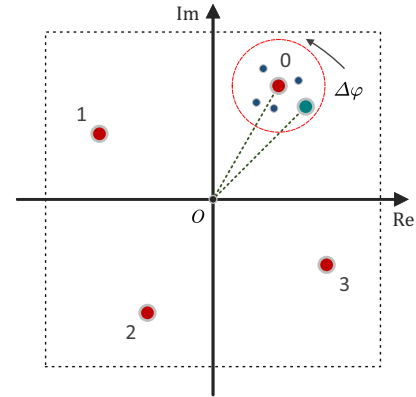}
    \caption{Certain phase deviation of constellation topology (QPSK).}
    \label{Figure: Phase Deviation}
\end{figure}

The analysis presented above also applies to flipping, as illustrated in Fig. \ref{Figure: Flipping}. Hence, we believe that rotation and flipping are intrinsically unified. However, we do not recommend using flipping in RF fingerprinting. The reason is that a radio signal affected by specific transmitter characteristics usually has its constellation topology deviated from standard morphology by a certain phase, as illustrated in Fig. \ref{Figure: Phase Deviation}. This deviation is part of RF fingerprints and varies from device to device, which cannot be corrected. The constellation topology becomes no longer strictly symmetric regarding both real and imaginary axes, making one symbol and all of its belonging sampling points cannot accurately transform into another, resulting in additional noise.

\vspace{10pt}

\noindent\textbf{Noise Adding} Huang \emph{et al.} \cite{huang2019data} also proposed adding Gaussian noise for signal data augmentation. However, we do not recommend using it in RF fingerprinting because such phase and amplitude distortions caused by RF fingerprints are usually very small and easily obscured by noise. In other words, RF fingerprints are relatively sensitive to noise.

\vspace{10pt}

\noindent\textbf{GAN-Based Generation} Considering that GANs can learn from data distribution to synthesize realistic samples \cite{wang2021generative}, some previous studies have discussed using GANs for signal data augmentation \cite{tang2018digital, patel2020data, lee2021uniqgan}. In practice, however, such GAN-based data augmentation usually performs poorly, especially when a radio signal is high-order modulated, affected by specific transmitter characteristics, or experienced channel fading. The main reason is that an underlying sampling distribution of radio signals is very complex in these cases and cannot be accurately modeled. Moreover, training such a GAN also requires a certain amount of labeled data, whereas a small dataset may not allow GANs to be well-optimized. The final performance could deteriorate further if many low-quality synthesized samples are introduced into training. Hence, we do not recommend using it in RF fingerprinting.

\subsection{Stochastic Permutation}
\begin{figure}[htb]
    \centering
    \includegraphics[scale=0.8125]{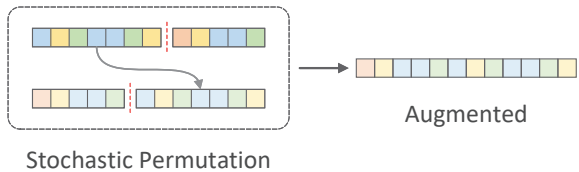}
    \caption{Illustration of stochastic permutation.}
    \label{Figure: Stochastic Permutation}
\end{figure}

\begin{figure}[htb]
    \centering
    \includegraphics[scale=1.35]{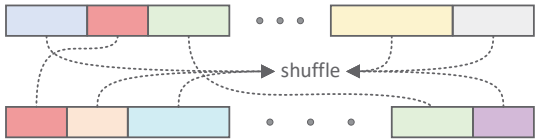}
    \caption{Illustration of stochastic permutation with more segments split.}
    \label{Figure: K Segmented Stochastic Permutation}
\end{figure}

Combined with specific domain knowledge, we also present another novel data augmentation for radio signals. The signal sample is a long sequence that consists of many complex-valued sampling points if its baseband form is considered, each of which can be considered drawn from an underlying sampling distribution of radio signals. In other words, these sampling points are identically distributed. This characteristic allows us to swap them arbitrarily. The simplest is to split a radio signal into two segments from an arbitrary middle position and then swap them, concatenated into a new sequence, as illustrated in Fig. \ref{Figure: Stochastic Permutation}. For clarity, we refer to this operation as ``stochastic permutation". The number of segments (denoted by $k$) can be further increased, termed as ``$k$-segmented stochastic permutation", as illustrated in Fig. \ref{Figure: K Segmented Stochastic Permutation}.

The sampling points in a signal sample could be mutually independent but are more likely to be partially correlated due to filtering or other similar factors. Considering shape filtering, to ensure symbol integrity as much as possible, we generally suggest $k \le \ell/\mathrm{sps}$, where $\ell$ denotes sample length and $\mathrm{sps}$ denotes oversampling ratio. Moreover, it should be pointed out that the effectiveness of stochastic permutation depends on the randomness of the emitted symbols. The effect is likely to weaken if a radio signal contains many continuous and repeated symbol fragments, even being invalid in some extreme cases, such as all zeros.

\subsection{Composite Data Augmentation}
As stated before, among existing data augmentations for radio signals, we have concluded that only rotation can play an effective role in RF fingerprinting. Besides, we also present another effective data augmentation operation, i.e., stochastic permutation. To maximize augmentation effects, we combine these two operations as a composite data augmentation scheme, which takes rotation as its primary transformation (randomly selecting one from a set of available rotation operations every time), followed by stochastic permutation. The whole procedure is expressed as a function $\operatorname{g}(\cdot)$.

\section{Methodology} \label{Section: Methodology}
Formally, we are provided with a radio dataset $\mathcal{D} = \mathcal{S} \cup \mathcal{U}$ collected from $\mathrm{C}$ wireless devices, where signals in $\mathcal{S}$ are labeled, i.e., $\mathcal{S} = \{(\boldsymbol{x}_i, \, y_i)\}_{i=1}^{\mathrm{M} \times \mathrm{C}}$, and those in $\mathcal{U}$ are not, i.e., $\mathcal{U} = \{\boldsymbol{x}_j\}_{j=1}^{\mathrm{N} \times \mathrm{C}}$, typically $\mathrm{M} \ll \mathrm{N}$. For clarity, we refer to a signal instance with its label as a signal example and those unlabeled ones as signal samples. The core problem is how to utilize $\mathcal{U}$ to help a given deep model $f_\theta$ learning on $\mathcal{S}$ for RF fingerprinting with improved generalization.

\subsection{Consistency-Based Regularization}
Deep learning employs backpropagation algorithms to optimize a target loss. The simplicity and efficiency of backpropagation algorithms for a great variety of loss functions make it attractive to add an unsupervised component to existing supervised objectives (e.g., cross-entropy), as follows.
\begin{equation}
    \mathcal{L} = \mathcal{L}_\mathrm{s} + \lambda \mathcal{L}_\mathrm{u}
\end{equation}
where $\mathcal{L}_\mathrm{s}$ represents a supervised objective, $\mathcal{L}_\mathrm{u}$ is an unsupervised component, and $\lambda$ is a penalty factor. This additional unsupervised term can be considered a form of regularization over unlabeled data, employed by virtually all deep semi-supervised learning methods \cite{van2020survey}. In particular, consistency-based regularization is well acclaimed due to its state-of-the-art (SOTA) performance \cite{oliver2018realistic}.

\begin{figure}[htb]
    \centering
    \includegraphics[scale=0.785]{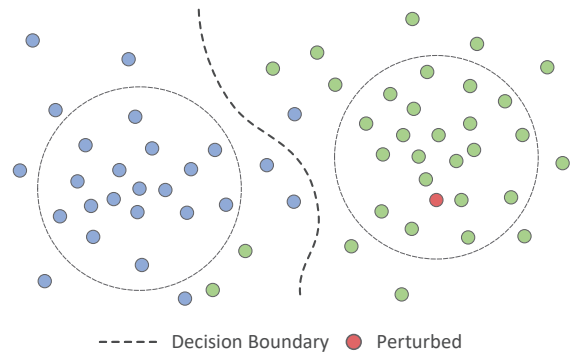}
    \caption{Consistency-based regularization with smoothness and clustering assumptions.}
    \label{Figure: Smoothness and low-density Assumption}
\end{figure}

As illustrated in Fig. \ref{Figure: Smoothness and low-density Assumption}, consistency-based regularization relies on smoothness and clustering assumptions. Concretely, when an input sample was applied with some realistic perturbation, its corresponding model prediction should not change significantly because these data points with distinct labels are separated by low-density regions, making it unlikely that one sample switches its class after being perturbed \cite{ouali2020overview}.

\vspace{10pt}

\noindent\textbf{Build Perturbation} It is not difficult to find that how to perturb unlabeled data is a key aspect of consistency-based regularization. Earlier work often introduces random noise for perturbation \cite{sajjadi2016regularization, samuli2017temporal}. More recently, it has been proved that using strong data augmentation for perturbation could obtain better results \cite{xie2019unsupervised, berthelot2019remixmatch}. Hence, we shall leverage our composite data augmentation to achieve perturbation. Notably, when $\operatorname{g}(\cdot)$ is used to perturb an unlabeled signal sample, $\mathrm{rot}_{0^\circ}$ should be discarded since it does not yield any perturbation effect.

The perturbation is then enhanced further with a simple weighted strategy. Let $\tilde{p}_t = f_\theta(\operatorname{g}(\boldsymbol{x}))$ be a prediction after being perturbed in epoch $t$ for $\boldsymbol{x} \in \mathcal{U}$. The prediction in epoch $t-1$ is always retained, denoted by $\tilde{p}_{t-1}$. The same sample during different training epochs (i.e., current and previous) corresponds to different perturbed outputs. They usually have different confidence levels even when predicted to be an identical class. Hence, we can combine $\tilde{p}_t$ and $\tilde{p}_{t-1}$ using a coefficient $\kappa$ to form a final perturbed output:
\begin{equation}
    \tilde{p} = \kappa \tilde{p}_{t-1} + (1-\kappa) \tilde{p}_t
\end{equation}

\vspace{10pt}

\noindent\textbf{How to Enforce Consistency} Another important aspect of consistency-based regularization is to enforce consistency, which encourages a model to have close predictions on an unlabeled sample and its perturbed versions. Such expected similarity can be computed using some popular distance measures like mean squared error (MSE), Kullback-Leiber (KL) divergence, and Jensen-Shannon (JS) divergence, which are not dependent on any true label. Inspired by FixMatch \cite{sohn2020fixmatch}, a newly recent SOTA algorithm for semi-supervised image classification, we incorporate pseudo-labeling into consistency-based regularization and then can measure this similarity directly with cross-entropy.

Pseudo-labeling is not competitive against other modern semi-supervised learning algorithms on its own, but using it as a part of these modern ones could usually produce better results. Given a signal sample $\boldsymbol{x} \in \mathcal{U}$ and its corresponding model prediction after being perturbed, i.e., $\tilde{p}$, we can convert $\tilde{p}$ into a hard label $\hat{y}$ (known as ``one-hot" encoding), and it is retained only when $\tilde{p}$ enjoys high confidence:
\begin{equation}
    \hat{y} = \mathop{\arg\mathop{\max}\limits_{c}}(\tilde{p}) \quad \text{if} \ \Omega_c(\tilde{p}) \ge \tau
\end{equation}
where $c \in \{1, \ldots, \mathrm{C}\}$ represents a possible class. The predicted class probability of $\tilde{p}$ on $c$ is $\Omega_c(\tilde{p})$, and $\tau$ is a preset threshold. Hence, we can obtain a certain number of signal samples with their pseudo-labels, which constitute a pseudo-labeled set, i.e., $\hat{\mathcal{S}} = \{(\boldsymbol{x}, \, \hat{y})\}$. Then, we can apply cross-entropy over such a pseudo-labeled set to enforce consistency.

\subsection{Training Pipeline}
The related components of our proposed method, including data augmentation and consistency-based regularization, have been elaborated, and we shall assemble them into an end-to-end training pipeline. To begin with, we leverage $\operatorname{g}(\cdot)$ to increase a limited amount of labeled data. The supervised loss is calculated using a standard cross-entropy loss, i.e.,
\begin{equation}
    \mathcal{L}_\mathrm{s} = \displaystyle \frac{1}{|\mathcal{S}|} \sum_{(\boldsymbol{x}, \, y) \in \mathcal{S}} {H(f_\theta(\operatorname{g}(\boldsymbol{x})), \; y)}
\end{equation}

Then, we perform consistency-based regularization using a large amount of unlabeled data. As noted, we can construct a pseudo-labeled set from these unlabeled samples. The unsupervised loss is also calculated using a standard cross-entropy loss over this pseudo-labeled set, i.e.,
\begin{equation}
    \mathcal{L}_\mathrm{u} = \displaystyle \frac{1}{|\hat{\mathcal{S}}|} \sum_{(\boldsymbol{x}, \, \hat{y}) \in \hat{\mathcal{S}}} {H(f_\theta(\boldsymbol{x}), \; \hat{y})}
\end{equation}

Considering the calculation order between $\mathcal{L}_\mathrm{u}$ and $\hat{\mathcal{S}}$, we reconstruct $\hat{\mathcal{S}}$ at the end of every epoch, used for calculating $\mathcal{L}_\mathrm{u}$ in the next epoch. The unsupervised loss $\mathcal{L}_\mathrm{u}$ is finally attached to the supervised loss $\mathcal{L}_\mathrm{s}$ as a regularization term, allowing $f_\theta$ to work in a semi-supervised fashion. Since both our supervised and unsupervised losses are calculated by cross-entropy, and there are usually few high confidence predictions at the training beginning, we neither need a warm-up procedure commonly used by other semi-supervised algorithms nor carefully tune a penalty factor. The supervised and unsupervised losses can be directly added. The full objective thus can be written as:
\begin{equation}
    \mathcal{L} = \mathcal{L}_\mathrm{s} + \mathcal{L}_\mathrm{u}
\end{equation}

For clarity, we summarize all these steps in Algorithm \ref{Algorithm: Semi-Supervised RF Fingerprinting}.

\begin{algorithm}
    \label{Algorithm: Semi-Supervised RF Fingerprinting}
    \caption{Semi-Supervised RF Fingerprinting}
    \SetAlgoLined
    \SetKwInOut{KIN}{Prepare}
    \KIN{Radio dataset $\mathcal{D} = \mathcal{S} \cup \mathcal{U}$, weighted coefficient $\kappa$, threshold $\tau$, model $f_\theta(\cdot)$, maximum number of iterative epochs $\mathrm{T}$.}
    \textbf{Initialize} $\hat{\mathcal{S}} = \varnothing$, $t = 1$, $\tilde{p}_0 = \boldsymbol{0}$. \\
    \While{$t \le \mathrm{T}$}{
        \vskip 0.125cm
        $
            \mathcal{L}_\mathrm{s} = \displaystyle \frac{1}{|\mathcal{S}|} \sum_{(\boldsymbol{x}, \, y) \in \mathcal{S}} {H(f_\theta(\operatorname{g}(\boldsymbol{x})), \; y)}
        $ \quad\textcolor[RGB]{192, 192, 192}{// Supervised} \\
        \vskip 0.125cm
        $
            \mathcal{L}_\mathrm{u} = \displaystyle \frac{1}{|\hat{\mathcal{S}}|} \sum_{(\boldsymbol{x}, \, \hat{y}) \in \hat{\mathcal{S}}} {H(f_\theta(\boldsymbol{x}), \; \hat{y})}
        $ \quad\textcolor[RGB]{192, 192, 192}{// Unsupervised} \\
        \vskip 0.125cm
        $
            \mathcal{L} = \mathcal{L}_\mathrm{s} + \mathcal{L}_\mathrm{u}
        $ \quad\textcolor[RGB]{192, 192, 192}{// Computed with mini-batch} \\
        \vskip 0.125cm
        \textbf{Optimize} The gradients with respect to $\mathcal{L}$ are computed by backpropagation and then update $\theta$ using gradient descent. \\
        \vskip 0.125cm
        \textbf{Pseudo-Labeling} The pseudo-labeled set needs to be reconstructed at each epoch ending: \\
        \vskip 0.125cm
        Reset $\hat{\mathcal{S}}$ as an empty set. \\
        \vskip 0.125cm
        \For{$\boldsymbol{x} \in \mathcal{U}$}{
            \vskip 0.125cm
            $\tilde{p}_t = f_\theta(\operatorname{g}(\boldsymbol{x}))$
            \quad\textcolor[RGB]{192, 192, 192}{// Record it for next epoch ...} \\
            \vskip 0.125cm
            $\tilde{p} = \kappa \tilde{p}_{t-1} + (1-\kappa) \tilde{p}_t$ \\
            \vskip 0.125cm
            \uIf{$\max(\tilde{p}) > \tau$}{
                \vskip 0.125cm
                \,$\hat{y} \coloneqq \mathop{\arg\mathop{\max}\limits_{c}}(\tilde{p})$ \\
                \vskip 0.125cm
                $\hat{\mathcal{S}} \coloneqq \hat{\mathcal{S}} \cup (\boldsymbol{x}, \, \hat{y})$ \\
            }
        }
    }
\end{algorithm}

It should be pointed out that our proposed method shares many similarities with FixMatch \cite{sohn2020fixmatch}. Both comprise a combination of two existing techniques: consistency-based regularization and pseudo-labeling. Note that FixMatch leverages two kinds of augmentations: ``weak" and ``strong". Specifically, FixMatch uses a weakly-augmented version of an unlabeled sample for prediction to yield a pseudo-label and enforce consistency against its strongly-augmented version. However, we no longer distinguish between ``strong" and ``weak" data augmentations and only consider single data augmentation, which is only used to perturb an unlabeled sample to get its pseudo-label, and then we enforce consistency against this sample itself, not its augmented version. In addition, we employ a weighting strategy of model weights to enhance perturbation further. Most importantly, data augmentation should be customized according to our respective fields. These similarities and differences suggest that our method can be viewed as a variant of FixMatch used for a specific field.

\section{Experiments and Results} \label{Section: Experiments and Results}
In this section, a series of experiments are conducted to evaluate our proposed method comprehensively.

\subsection{Data Preparation}
Note that our experiments include many ablation studies and necessary comparisons with existing methods. In ablation experiments, we consider using communication simulation to generate a radio dataset since it can facilitate precise control of various experimental conditions and exclude other possible interferences, which allows us to observe relevant phenomena and trends more clearly and stably. In comparison experiments, we adopt both simulated and real-world datasets, which aims to make related results more sufficient and convincing. The details about our datasets are described as follows.

\vspace{10pt}

\noindent\textbf{Simulated Dataset} The radio signal may be a function of amplitude, phase, frequency, or combination of these \cite{goldsmith2005wireless}. In general, we can represent it as
\begin{equation}
    r(t) = \mathrm{Re} \left\{ a(t) e^{2 \pi f_c t + \varphi(t)} \right\} + n(t)
\end{equation}
where $f_c$ denotes carrier frequency, $a(t)$ and $\varphi(t)$ are modulated envelope and phase, respectively, and $n(t)$ denotes channel noise. The inherent imperfections of hardware, such as clock jitter, digital-to-analog (D/A) converters, mixers or local frequency synthesizers, and power amplifiers' nonlinearity, could cause a certain degree of signal distortions, whose comprehensive impacts can be summarized as two non-linear transformations: amplitude function $A(\cdot)$ and phase function $\varPhi(\cdot)$. The radio signal with specific RF fingerprints thus can be rewritten as
\begin{equation}
    r(t) = \mathrm{Re} \left\{ A\left[a(t)\right] e^{2 \pi f_c t + \varphi(t) + \varPhi\left[a(t)\right]} \right\} + n(t)
\end{equation}

\begin{figure}[htb]
    \centering
    \includegraphics[scale=0.6]{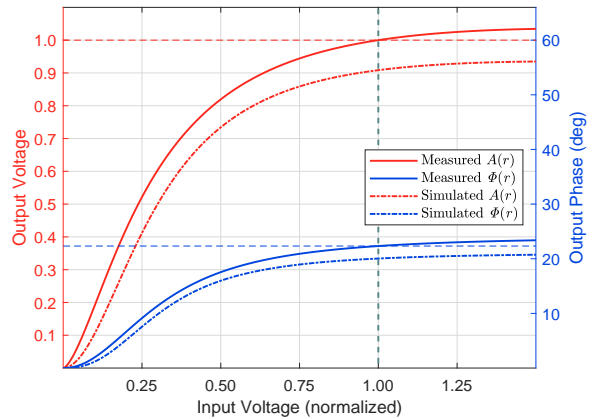}
    \caption{Simulation of RF fingerprints via adjusting a measured power amplification characteristic curve.}
    \label{Figure: Power Amplification Characteristic}
\end{figure}

The nonlinearity of power amplifiers is often considered an important factor to form RF fingerprints \cite{hanna2019deep}. Aghasi \emph{et al.} \cite{aghasi2007modified} proposed a power amplifier model for solid-state power amplifiers (SSPA) where $A(r)$ is a function of input amplitude $r$, representing AM/AM conversion and $\varPhi(r)$, also a function of input amplitude $r$, representing AM/PM conversion as:
\begin{equation}
    \left\{
    \begin{aligned}
        A(r)       & = \displaystyle \frac{\alpha_1 r^{\alpha_2} + \alpha_3 r^{\alpha_2 + 1}}{1 + \alpha_4 r^{\alpha_2 + 1}} \\
        \varPhi(r) & = \displaystyle \frac{\beta_1 r^{\beta_2} + \beta_3 r^{\beta_2 + 1}}{1 + \beta_4 r^{\beta_2 + 1}}
    \end{aligned} \right.
\end{equation}
Note that different coefficient choices for $\boldsymbol{\alpha}$ and $\boldsymbol{\beta}$ can produce different amplification characteristic curves to represent different devices. As illustrated in Fig. \ref{Figure: Power Amplification Characteristic}, we consider a set of coefficients measured from a GaAs FET power amplifier and then apply a small random offset to produce multiple simulated devices. There are a total of $10$ devices simulated, as listed in Table \ref{Table: Simulation with Non-Linear Power Amplification}, and their RF fingerprint differences are very small.

\begin{table*}[htb]
    \renewcommand\arraystretch{1.15}
    \centering
    \caption{Simulation with Non-Linear Power Amplification Model}
    \label{Table: Simulation with Non-Linear Power Amplification}
    \resizebox{0.7\textwidth}{!}{
        \begin{tabular}{ccc}
            \toprule
            \textbf{No.} & \textbf{Amplitude ($\alpha$)}           & \textbf{Phase ($\beta$)}                \\
            \midrule
            \specialrule{0em}{1pt}{0pt}
            \midrule
            XMTR0        & $(10.2598,\,1.9926,\,-0.2782,\,9.5297)$ & $(6.0838,\,1.3190,\,-0.0375,\,16.2325)$ \\
            XMTR1        & $(10.7344,\,2.0668,\,-0.5015,\,9.5297)$ & $(6.3304,\,1.3058,\,-0.0348,\,16.2325)$ \\
            XMTR2        & $(11.6849,\,2.0193,\,-0.6689,\,9.5297)$ & $(6.7758,\,2.0689,\,-0.0280,\,16.2325)$ \\
            XMTR3        & $(10.2963,\,1.7932,\,-0.2929,\,9.5297)$ & $(6.1256,\,1.4660,\,-0.0297,\,16.2325)$ \\
            XMTR4        & $(11.3625,\,2.0100,\,-0.4304,\,9.5297)$ & $(6.6729,\,2.2441,\,-0.0168,\,16.2325)$ \\
            XMTR5        & $(11.4996,\,2.0766,\,-0.5835,\,9.5297)$ & $(6.7440,\,2.9490,\,-0.0454,\,16.2325)$ \\
            XMTR6        & $(10.5223,\,1.7999,\,-0.5658,\,9.5297)$ & $(6.4241,\,1.4531,\,-0.0425,\,16.2325)$ \\
            XMTR7        & $(10.4870,\,1.8997,\,-0.4515,\,9.5297)$ & $(6.4135,\,1.4193,\,-0.0305,\,16.2325)$ \\
            XMTR8        & $(11.3525,\,2.2360,\,-0.2442,\,9.5297)$ & $(6.9513,\,2.1135,\,-0.0366,\,16.2325)$ \\
            XMTR9        & $(10.0237,\,1.9307,\,-0.4582,\,9.5297)$ & $(6.0633,\,2.4454,\,-0.0363,\,16.2325)$ \\
            \bottomrule
        \end{tabular}}
\end{table*}

The other important parameters for communication simulation are depicted as follows. The emitted message is completely random and adopts QPSK modulation. The square-root raised cosine FIR filter with a rolloff factor of $0.35$ is used for pulse-shaping. The additive white Gaussian noise (AWGN) channel is considered. The signal-to-noise ratio (SNR) is set to $18$ dB. Each signal sample has $1024$ sampling points with $8 \times$ oversampling. There are a total of $10000$ samples per device generated for each trial. The signal data is randomly divided into a training set, validation set, and test set in a proportion of $3:1:1$. The training set thus has $6000$ samples per device, from which we can take a portion for supervised learning with a limited number of examples or further distinguish between labeled and unlabeled data for semi-supervised learning.

\vspace{10pt}

\noindent\textbf{Real-World Dataset} To be more convincing, we consider using an open-source RF dataset for comparison, WIDEFT \cite{siddik2021wideft}, available on \href{https://zenodo.org/record/4116383}{Zenodo}. This dataset is collected from $138$ real-world devices (e.g., smartphones, headsets, routers), spanning $79$ unique models, some of which (e.g., smartphones) are capable of more than one wireless type (e.g., Bluetooth/WiFi) and radio modes ($2.4 \text{/} 5$ GHz WiFi), leading to a total of $147$ data captures made. Each capture contains a set of $100$ bursts, each of which is complete that consists of ON transient, steady-state portion, and OFF transient, and includes $5000$ sampling points before and after. The steady-state portion is long enough to be sliced into multiple samples, and we randomly slice $50$ samples of length $1024$ from each burst. Table \ref{Table: Model List} shows that all Apple Inc products with $2.4$ GHz WiFi in WIDEFT are selected to support our experiments. The signal data is divided into a training set, validation set, and test set in a proportion of $3:1:1$ (counted in bursts). There are $60$ bursts per device available for training, from which we can further distinguish between labeled and unlabeled data for semi-supervised learning.

\begin{table}[htb]
    \centering
    \caption{Model List of Apple Inc's $2.4$ GHz WiFi products in WIDEFT}
    \label{Table: Model List}
    \begin{tabular}{ccc}
        \toprule
        \textbf{Model} & \textbf{Num.} & \textbf{Description}                \\
        \midrule
        \specialrule{0em}{1pt}{0pt}
        \midrule
        A1241          & 1             & Smartphone, iPhone 3G               \\
        A1349          & 1             & Smartphone, iPhone 4                \\
        A1367          & 3             & Media Player, iPod touch 4th Gen    \\
        A1387          & 1             & Smartphone, iPhone 4s               \\
        A1432          & 1             & Tablet, iPad mini                   \\
        A1453          & 1             & Smartphone, iPhone 5s               \\
        A1534          & 1             & Notebook, MacBook                   \\
        A1660          & 1             & Smartphone, iPhone 7                \\
        A1708          & 1             & Notebook, MacBook Pro 13"           \\
        A1860          & 1             & Smartwatch, Watch Series 3          \\
        A1876          & 1             & Tablet, iPad Pro 12.9 3th Gen       \\
        A1905          & 1             & Smartphone, iPhone 8                \\
        A1920          & 1             & Smartphone, iPhone XS               \\
        A1975          & 1             & Smartwatch, Watch Series 4 Cellualr \\
        A1978          & 1             & Smartwatch, Watch Series 4          \\
        A2197          & 1             & Tablet, iPad 7th Gen                \\
        \bottomrule
    \end{tabular}
\end{table}


\begin{figure}[htb]
    \centering
    \includegraphics[scale=0.765]{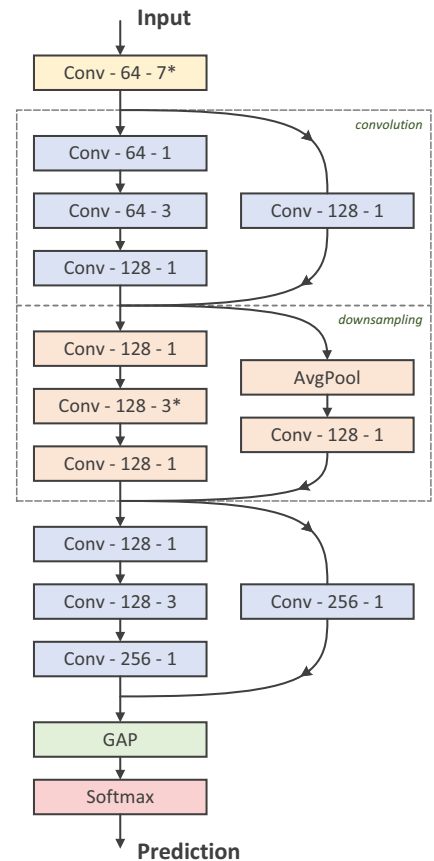}
    \caption{The structure of ResNet we designed for RF fingerprinting. The layer label ``Conv-$k$-$s$'' indicates that this convolutional layer has $k$ kernels of size $s$. The convolutions of size $3$ are all separable. The asterisk (*) indicates that this layer enjoys a stride of size $2$. Batch normalization (BN) \cite{ioffe2015batch} has been performed after each convolution, followed by activation using rectified linear units (ReLU) \cite{nair2010rectified}.}
    \label{Figure: Network}
\end{figure}

\subsection{Implementation and Training Details}
This work essentially provides a semi-supervised training pipeline for RF fingerprinting. The target network to be trained ($f_\theta$) could be implemented with any network architecture as long as it can be used for classification. Previous literature has discussed many network architectures for RF fingerprinting. The most commonly used are still a variety of convolutional architectures. Popular convolutional architectures and their variants include VGG \cite{simonyan2014very}, Inception \cite{szegedy2015going}, ResNet \cite{he2016deep}, etc.

In order to achieve satisfactory performance, including supervised and semi-supervised performance, we have carefully designed a ResNet. The network starts from a convolutional layer, followed by a series of alternately stacked convolution and downsampling blocks, and ends with a classification layer. The initial convolutional layer is used as an input stem, and it has $64$ kernels of size $7$, with a stride of $2$. The convolution and downsampling blocks are both implemented as residual blocks, and their designs mainly refer to ResNet-B and ResNet-D mentioned in \cite{he2019bag}. The last classification layer performs global average pooling (GAP) and then yields a prediction using softmax. The number of residual blocks could be varied to change modeling capability. The ResNet we used has two contiguous convolution blocks and employs a downsampling block as their transition. The detailed structure is illustrated in Fig. \ref{Figure: Network}.

The model is built with TensorFlow \cite{tensorflow2015} and then trained on a single NVIDIA RTX 2080S GPU utilizing an Adam \cite{kingma2014adam} optimizer for $230$ epochs. The batch size can affect stability and convergence during training, and it cannot be too large when training with a limited number of examples, as this may lead to smooth gradient changes. The initial learning rate for supervised learning is set to $0.001$. The initial learning rate for semi-supervised learning is reduced to $3 \times 10^{-4}$, with better stability. The recognition accuracy is used as a performance metric and calculated based on $20$ trials.

\subsection{Ablation Study of Data Augmentation}
\begin{figure}[htb]
    \centering
    \includegraphics[scale=0.6]{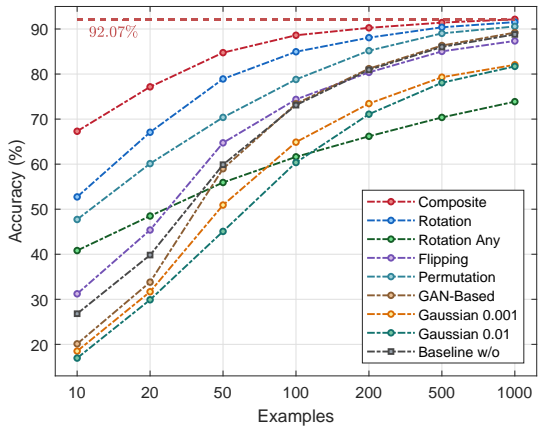}
    \caption{Comparison among different signal data augmentations. The overall recognition accuracy with fully supervised learning is $92.07\%$, achieved with $5000$ examples per device available for training.}
    \label{Figure: Data Augmentation}
\end{figure}

The proposed method for semi-supervised RF fingerprinting largely relies on data augmentation. Although a variety of data augmentations for radio signals have been comprehensively analyzed, we still report their actual performance in Fig. \ref{Figure: Data Augmentation}. It can be seen that our composite data augmentation is far superior to all data augmentations in a non-composite form, especially for cases with an extremely small number of examples available for training, which fully proves that this composite design is effective.

\begin{figure}[htb]
    \centering
    \includegraphics[scale=0.6]{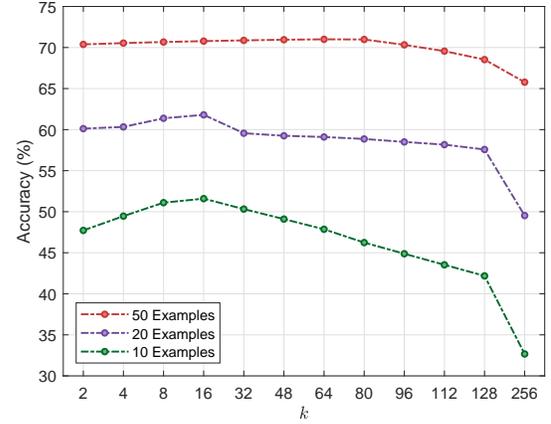}
    \caption{Effects of stochastic permutation when taking different $k$.}
    \label{Figure: Study of K Segments}
\end{figure}

Besides, we can see that RF fingerprinting does not benefit from adding Gaussian noise and instead becomes worse. Likewise, using GANs has a poor effect. As expected, flipping almost does not work in RF fingerprinting, whereas rotation works reasonably well, but such rotation should be operated only at specific angles, e.g., $0^\circ$, $90^\circ$, $180^\circ$, $270^\circ$ for QPSK. In particular, rotating at arbitrary angles and flipping can bring a little improvement in those cases with an extremely small number of examples, and we infer that they might work as regularization. Furthermore, we can see that stochastic permutation ($k = 2$) also behaves very well, and its augmentation effect is second to rotation among all non-composite data augmentations. Note that stochastic permutation is likely to perform better if we further increase $k$, as shown in Fig. \ref{Figure: Study of K Segments}. The above results are basically consistent with our previous analysis of data augmentation.

\subsection{Ablation Study of Hyper-Parameters}
The proposed method for semi-supervised RF fingerprinting has two hyper-parameters, i.e., a coefficient for weighting to enhance perturbation ($\kappa$) and a threshold for pseudo-labeling ($\tau$). Intuitively, we argue that $\kappa$ should be between $0$ and $0.5$. The reason is that the model weights of the current epoch are undoubtedly more relevant to the predictions of the current epoch. In theory, only $0.5 \le \tau < 1.0$ is meaningful. There would always be only a small number of samples labeled if we set $\tau$ relatively high, and then consistency-based regularization hardly works. Conversely, plenty of samples might be mislabeled if we set $\tau$ too small, introducing a high estimation bias.

\begin{figure*}[htb]
    \centering
    \subfigcapskip=-0.5em
    \subfigure[Trained with $\mathrm{M} = 10$ \& $\mathrm{N} = 1000$]
    {
        \centering
        \includegraphics[width=0.26\textwidth]{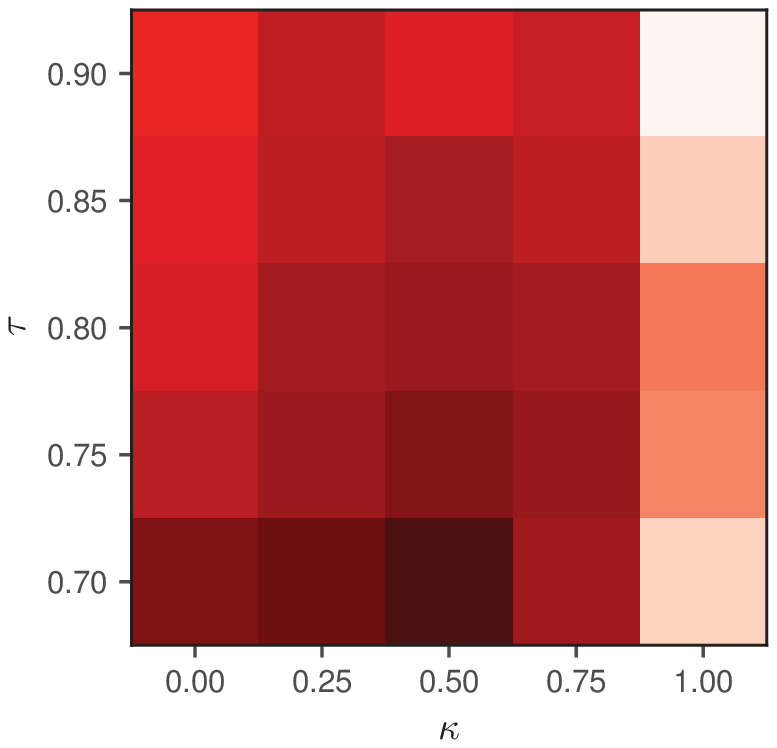}
    }
    \hspace{0.85em}
    \subfigure[Trained with $\mathrm{M} = 20$ \& $\mathrm{N} = 1000$]
    {
        \centering
        \includegraphics[width=0.26\textwidth]{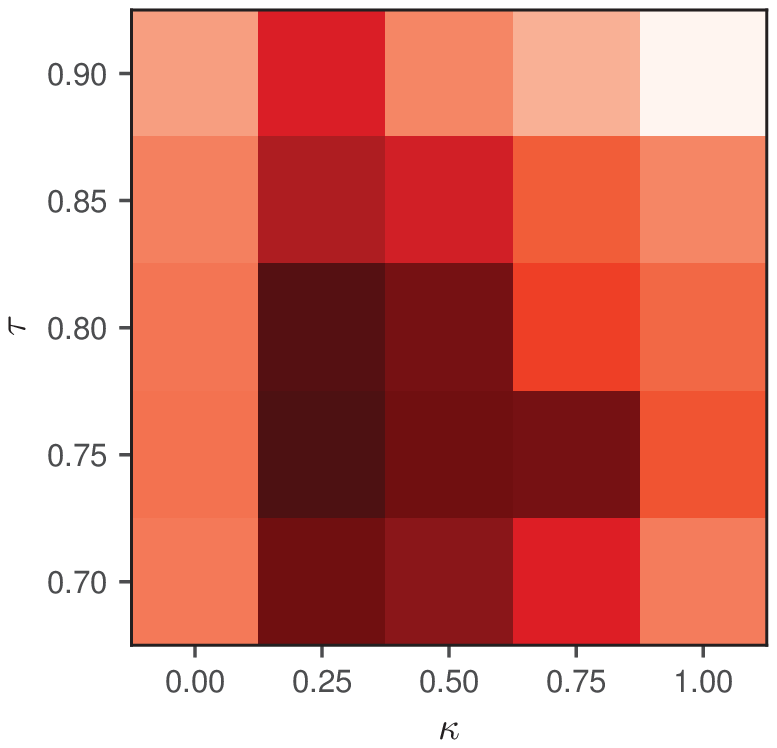}
    }
    \hspace{0.85em}
    \subfigure[Trained with $\mathrm{M} = 50$ \& $\mathrm{N} = 1000$]
    {
        \centering
        \includegraphics[width=0.26\textwidth]{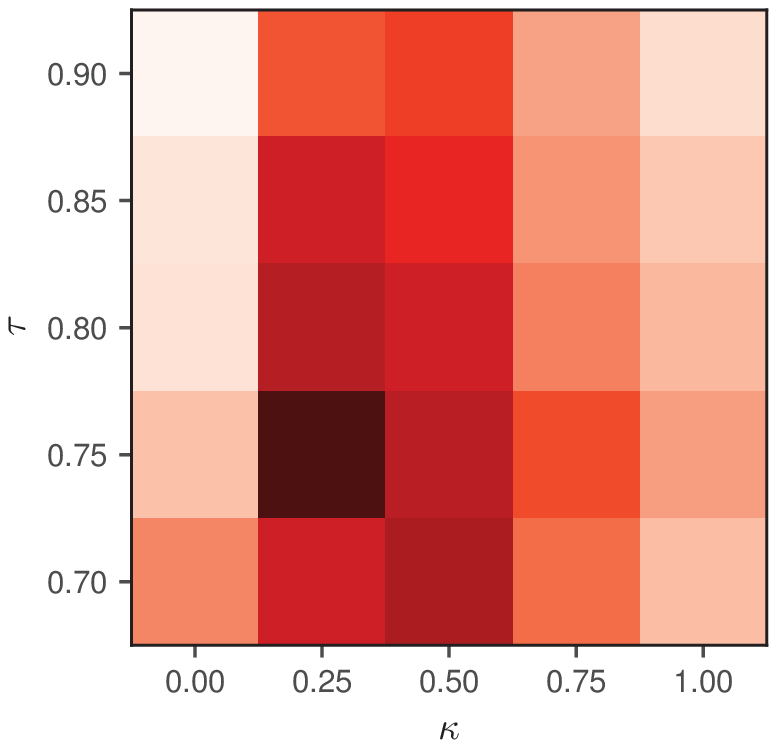}
    }
    \caption{Grid search for hyper-parameters, $\kappa$ and $\tau$. The darker represents higher recognition accuracy.}
    \label{Figure: Grid Search}
\end{figure*}

To tune $\kappa$ and $\tau$ reasonably, we conduct a two-dimensional grid search when $\mathrm{M} = 10, \, 20, \, 50$, and let $\mathrm{N} = 1000$, as shown in Fig. \ref{Figure: Grid Search}. The case of $\mathrm{M} = 10$ achieves its best accuracy at $\kappa = 0.50$ and $\tau = 0.70$, while both $\mathrm{M} = 20$ and $\mathrm{M} = 50$ achieve their best accuracies at $\kappa = 0.25$ and $\tau = 0.75$, from which we can conclude a rough trend that $\kappa$ should be decreased appropriately and $\tau$ should be raised slightly, when more examples are available. The configuration for $\kappa$ and $\tau$ might be different on different datasets, but we can often leverage their variation trends to obtain a relatively optimal choice.

\begin{figure}[htb]
    \centering
    \includegraphics[scale=0.6]{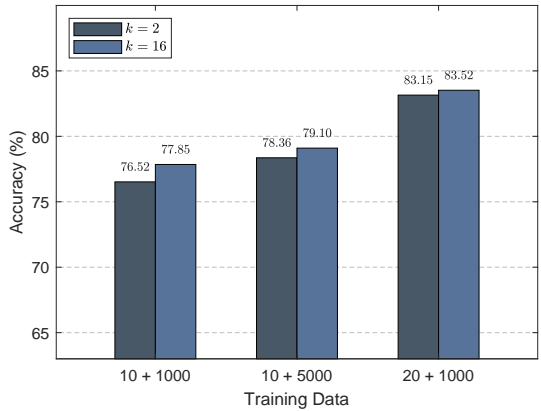}
    \caption{Increasing $k$ to improve semi-supervised performance.}
    \label{Figure: Semi-Supervised Learning with Different K}
\end{figure}

Besides, as demonstrated earlier, stochastic permutation can achieve better augmentation effects with relatively more segments split, i.e., appropriately increasing $k$, which can also impact our semi-supervised performance correspondingly. Consequently, $k$ can also be treated as a hyper-parameter in a certain sense. As shown in Fig. \ref{Figure: Semi-Supervised Learning with Different K}, our proposed method for semi-supervised RF fingerprinting has its recognition accuracy slightly improved when $k$ is appropriately increased ($\mathrm{M} = 10$ or $20$, $k$ increased from $2$ to $16$). Note that this improvement is relatively limited and cannot be comparable to what performs in supervised learning since this performance improvement space has already been largely exploited with semi-supervised learning itself. In contrast to such a little performance improvement, we might pay more attention to another benefit of appropriately increasing $k$. It is acknowledged that training stability could also be a challenge when training with very limited labeled data. For example, given $k = 2$, there are $3 \sim 7$ training failures in $20$ trials when $\mathrm{M} = 10$. Training failure refers to such cases that training loss becomes NaN too early or optimization falls into some bad minima, resulting in bad generalization. This issue can be significantly improved when increasing $k$ to $16$, where it only happens less than $3$ training failures in $20$ trials.

\subsection{Ablation Study of Network Complexity}
\begin{figure}[htb]
    \centering
    \includegraphics[scale=0.6]{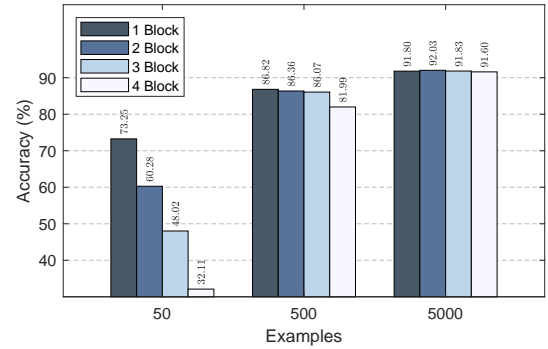}
    \caption{Supervised performance of our ResNet vs different numbers of convolution blocks stacked.}
    \label{Figure: Network Design for Supervised Learning}
\end{figure}

Network design is also essential for deep semi-supervised learning. It is acknowledged that a more complex neural network often implies better modeling capability. The complexity of a neural network largely depends on its width and depth. The network width has been restricted by GPU memory size. The general consideration is to increase depth, but a deeper neural network requires more data to be effectively trained. Otherwise, it is easy to occur overfitting. According to this principle, we often should consider a relatively simple network when training with a limited number of examples. The ResNet we designed can adjust its modeling capability with different numbers of convolution blocks stacked. As indicated in Fig. \ref{Figure: Network Design for Supervised Learning}, in supervised learning, our ResNet performs better when stacked by $1$ convolution block than $2$, $3$, and $4$ convolution blocks, especially with fewer examples. Shall we just implement it with only $1$ convolution block?

\begin{figure}[htb]
    \centering
    \includegraphics[scale=0.6]{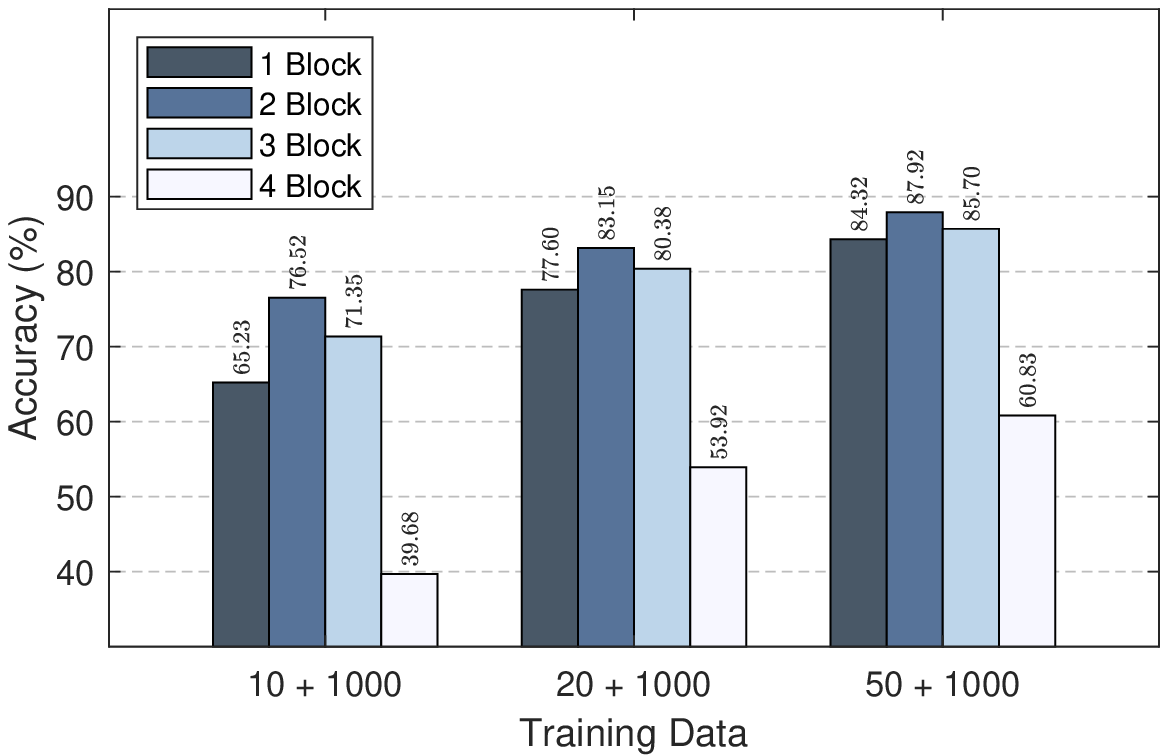}
    \caption{Semi-supervised performance of our ResNet vs different numbers of convolution blocks stacked.}
    \label{Figure: Network Design for Semi-Supervised Learning}
\end{figure}

\begin{table*}[htb]
    \renewcommand\arraystretch{1.15}
    \centering
    \caption{Comparsion on Simulated Dataset}
    \label{Table: Comparsion with Other Semi-Supervised Algorithms 1}
    \resizebox{0.96\textwidth}{!}{
        \begin{threeparttable}
            \begin{tabular}{cccccccccc}
                \toprule
                \multirow{2}{*}[-1.0ex]{\textbf{Method}}
                 & \multicolumn{6}{c}{$\boldsymbol{10}$}
                 & $\boldsymbol{20}$
                 & $\boldsymbol{50}$
                 & $\boldsymbol{180}$                    \\
                \cmidrule(lr){2-7} \cmidrule(lr){8-8} \cmidrule(lr){9-9} \cmidrule(lr){10-10}
                 & $\boldsymbol{100}$
                 & $\boldsymbol{200}$
                 & $\boldsymbol{500}$
                 & $\boldsymbol{1000}$
                 & $\boldsymbol{2000}$
                 & $\boldsymbol{5000}$
                 & $\boldsymbol{1000}$
                 & $\boldsymbol{1000}$
                 & $\boldsymbol{1800}$                   \\
                \midrule
                \specialrule{0em}{0.8pt}{0pt}
                \midrule
                Proposal
                 & $71.53 \pm 1.32$
                 & $73.06 \pm 1.30$
                 & $74.82 \pm 1.09$
                 & $76.52 \pm 0.65$
                 & $78.23 \pm 0.79$
                 & $78.36 \pm 0.69$
                 & $83.15 \pm 0.57$
                 & $87.92 \pm 0.33$
                 & $91.30 \pm 0.29$                      \\
                \midrule
                FixMatch
                 & $53.68 \pm 1.37$
                 & $67.39 \pm 1.25$
                 & $68.53 \pm 1.19$
                 & $68.72 \pm 1.16$
                 & $69.03 \pm 1.07$
                 & $69.30 \pm 1.03$
                 & $78.33 \pm 0.89$
                 & $85.37 \pm 0.70$
                 & $88.58 \pm 0.53$                      \\
                SSRCNN
                 & $17.68 \pm 0.39$
                 & $18.37 \pm 0.88$
                 & $18.74 \pm 0.50$
                 & $18.75 \pm 0.96$
                 & $17.86 \pm 1.15$
                 & $18.75 \pm 0.69$
                 & $31.30 \pm 1.39$
                 & $53.01 \pm 1.38$
                 & $78.20 \pm 0.76$                      \\
                E3SGAN
                 & $13.83 \pm 1.09$
                 & $14.87 \pm 1.30$
                 & $14.38 \pm 1.29$
                 & $14.10 \pm 1.24$
                 & $14.40 \pm 1.00$
                 & $15.19 \pm 1.52$
                 & $19.82 \pm 1.33$
                 & $21.38 \pm 0.88$
                 & $25.97 \pm 1.26$                      \\
                \bottomrule
            \end{tabular}
            \begin{tablenotes}
                \footnotesize
                \item The table headers like ``$\boldsymbol{10}$ \& $\boldsymbol{100}$'' means $\mathrm{M} = 10$ and $\mathrm{N} = 100$. The ``weak" and ``strong" data augmentations used by FixMatch are specified as rotation at specific angles and our composite data augmentation, respectively, which aims to adapt for RF fingerprinting.
            \end{tablenotes}
        \end{threeparttable}}
\end{table*}

\begin{table}[htb]
    \renewcommand\arraystretch{1.15}
    \centering
    \caption{Comparsion on Real-World Dataset, WIDEFT}
    \label{Table: Comparsion with Other Semi-Supervised Algorithms 2}
    \resizebox{0.45\textwidth}{!}{
        \begin{threeparttable}
            \begin{tabular}{cccc}
                \toprule
                \textbf{Method}
                 & $\boldsymbol{1} + \boldsymbol{10}$ & $\boldsymbol{2} + \boldsymbol{20}$ & $\boldsymbol{5} + \boldsymbol{50}$ \\
                \midrule
                \specialrule{0em}{0.8pt}{0pt}
                \midrule
                Proposal
                 & $86.05 \pm 0.37$
                 & $91.73 \pm 0.30$
                 & $93.82 \pm 0.25$
                \\
                \midrule
                FixMatch
                 & $81.76 \pm 0.53$
                 & $89.35 \pm 0.38$
                 & $93.25 \pm 0.32$
                \\
                SSRCNN
                 & $49.50 \pm 1.19$
                 & $75.67 \pm 0.98$
                 & $89.10 \pm 0.73$
                \\
                E3SGAN
                 & $16.85 \pm 1.52$
                 & $23.60 \pm 1.38$
                 & $37.08 \pm 1.09$
                \\
                \bottomrule
            \end{tabular}
            \begin{tablenotes}
                \footnotesize
                \item The table headers like ``$\boldsymbol{1} + \boldsymbol{10}$'' means $\mathrm{M} = 1$ and $\mathrm{N} = 10$.
            \end{tablenotes}
        \end{threeparttable}}
\end{table}

However, we can see from Fig. \ref{Figure: Network Design for Semi-Supervised Learning} that in semi-supervised learning, stacking only $1$ convolution block is far inferior to stacking $2$ or $3$ convolution blocks. And we infer that it is because consistency-based regularization must enforce deep neural networks to learn an intrinsic relation between unlabeled samples and their perturbed versions, which may require stronger modeling capability. The semi-supervised performance deteriorates again and even becomes unstable if we keep stacking more, still due to overfitting. In other words, we need to find a balance. Finally, we decide to stack $2$ convolution blocks as a relatively optimal configuration for semi-supervised learning.

\subsection{Comparsion with Other Semi-Supervised Algorithms}
The proposed method is compared with other competing ones, including two recent works on communication signal recognition using deep semi-supervised learning: SSRCNN \cite{dong2021ssrcnn} and E3SGAN \cite{zhou2020generative}. Since ours bears many resemblances to FixMatch \cite{sohn2020fixmatch}, we also compare with it but replace those image data augmentations originally used by FixMatch with our signal data augmentations.

To begin with, we investigate their performance on simulated datasets. The results are given in Table \ref{Table: Comparsion with Other Semi-Supervised Algorithms 1}. It can be seen that our proposed method is far superior to SSRCNN and E3SGAN. The reason largely lies in that these two do not consider any data augmentation, which indicates that data augmentation indeed plays an important role in semi-supervised learning. And ours also performs better than FixMatch, especially when labeled samples are very limited. Note that ours can achieve an overall recognition accuracy of more than $91.30\%$ at $\mathrm{M} = 180$ and $\mathrm{N} = 1800$, almost close to fully supervised learning (i.e., $92.07\%$, achieved with $5000$ signal examples per device available for training). The demand for labeled data is greatly reduced.

To be more convincing, we also employ a real-world RF dataset (i.e., WIDEFT) for comparison. The results are given in Table \ref{Table: Comparsion with Other Semi-Supervised Algorithms 2}. It can be seen that our proposed method still outperforms SSRCNN, E3SGAN, and FixMatch. The experimental results on both simulated and real-world RF datasets indicate that our proposed method is quite effective and reaches SOTA for semi-supervised RF fingerprinting.

\section{Conclusion} \label{Section: Conclusion}
This paper has introduced deep semi-supervised learning for RF fingerprinting, which largely relies on a composite data augmentation technique for radio signals in conjunction with an effective combination of consistency-based regularization and pseudo-labeling. The experimental results on both simulated and real-world datasets indicate that our proposed method for semi-supervised RF fingerprinting significantly outperforms competing ones. In future work, we would like to combine our proposed method with open-set recognition to increase practicability further.

\bibliographystyle{IEEEtran}
\bibliography{references}

\begin{thebibliography}{10}
\providecommand{\url}[1]{#1}
\csname url@samestyle\endcsname
\providecommand{\newblock}{\relax}
\providecommand{\bibinfo}[2]{#2}
\providecommand{\BIBentrySTDinterwordspacing}{\spaceskip=0pt\relax}
\providecommand{\BIBentryALTinterwordstretchfactor}{4}
\providecommand{\BIBentryALTinterwordspacing}{\spaceskip=\fontdimen2\font plus
\BIBentryALTinterwordstretchfactor\fontdimen3\font minus
  \fontdimen4\font\relax}
\providecommand{\BIBforeignlanguage}[2]{{%
\expandafter\ifx\csname l@#1\endcsname\relax
\typeout{** WARNING: IEEEtran.bst: No hyphenation pattern has been}%
\typeout{** loaded for the language `#1'. Using the pattern for}%
\typeout{** the default language instead.}%
\else
\language=\csname l@#1\endcsname
\fi
#2}}
\providecommand{\BIBdecl}{\relax}
\BIBdecl

\bibitem{zou2016survey}
Y.~Zou, J.~Zhu, X.~Wang, and L.~Hanzo, ``A survey on wireless security:
  Technical challenges, recent advances, and future trends,'' \emph{Proceedings
  of the IEEE}, vol. 104, no.~9, pp. 1727--1765, 2016.

\bibitem{hamamreh2018classifications}
J.~M. Hamamreh, H.~M. Furqan, and H.~Arslan, ``Classifications and applications
  of physical layer security techniques for confidentiality: A comprehensive
  survey,'' \emph{IEEE Communications Surveys \& Tutorials}, vol.~21, no.~2,
  pp. 1773--1828, 2018.

\bibitem{xu2015device}
Q.~Xu, R.~Zheng, W.~Saad, and Z.~Han, ``Device fingerprinting in wireless
  networks: Challenges and opportunities,'' \emph{IEEE Communications Surveys
  \& Tutorials}, vol.~18, no.~1, pp. 94--104, 2015.

\bibitem{soltanieh2020review}
N.~Soltanieh, Y.~Norouzi, Y.~Yang, and N.~C. Karmakar, ``A review of radio
  frequency fingerprinting techniques,'' \emph{IEEE Journal of Radio Frequency
  Identification}, vol.~4, no.~3, pp. 222--233, 2020.

\bibitem{lecun2015deep}
Y.~LeCun, Y.~Bengio, and G.~Hinton, ``Deep learning,'' \emph{nature}, vol. 521,
  no. 7553, pp. 436--444, 2015.

\bibitem{jian2020deep}
T.~Jian, B.~C. Rendon, E.~Ojuba, N.~Soltani, Z.~Wang, K.~Sankhe, A.~Gritsenko,
  J.~Dy, K.~Chowdhury, and S.~Ioannidis, ``Deep learning for rf fingerprinting:
  A massive experimental study,'' \emph{IEEE Internet of Things Magazine},
  vol.~3, no.~1, pp. 50--57, 2020.

\bibitem{o2016convolutional}
T.~J. O'Shea, J.~Corgan, and T.~C. Clancy, ``Convolutional radio modulation
  recognition networks,'' in \emph{International conference on engineering
  applications of neural networks}.\hskip 1em plus 0.5em minus 0.4em\relax
  Springer, 2016, pp. 213--226.

\bibitem{mao2018deep}
Q.~Mao, F.~Hu, and Q.~Hao, ``Deep learning for intelligent wireless networks: A
  comprehensive survey,'' \emph{IEEE Communications Surveys \& Tutorials},
  vol.~20, no.~4, pp. 2595--2621, 2018.

\bibitem{sun2019application}
Y.~Sun, M.~Peng, Y.~Zhou \emph{et~al.}, ``Application of machine learning in
  wireless networks: Key techniques and open issues,'' \emph{IEEE
  Communications Surveys \& Tutorials}, vol.~21, no.~4, pp. 3072--3108, 2019.

\bibitem{pham2021intelligent}
Q.-V. Pham, N.~T. Nguyen, T.~Huynh-The \emph{et~al.}, ``Intelligent radio
  signal processing: A survey,'' \emph{IEEE Access}, 2021.

\bibitem{riyaz2018deep}
S.~Riyaz, K.~Sankhe, S.~Ioannidis, and K.~Chowdhury, ``Deep learning
  convolutional neural networks for radio identification,'' \emph{IEEE
  Communications Magazine}, vol.~56, no.~9, pp. 146--152, 2018.

\bibitem{peng2019deep}
L.~Peng, J.~Zhang, M.~Liu, and A.~Hu, ``Deep learning based rf fingerprint
  identification using differential constellation trace figure,'' \emph{IEEE
  Transactions on Vehicular Technology}, vol.~69, no.~1, pp. 1091--1095, 2019.

\bibitem{baldini2019assessment}
G.~Baldini and R.~Giuliani, ``An assessment of the impact of wireless
  interferences on iot emitter identification using time frequency
  representations and cnn,'' in \emph{2019 Global IoT Summit (GIoTS)}.\hskip
  1em plus 0.5em minus 0.4em\relax IEEE, 2019, pp. 1--6.

\bibitem{yu2019robust}
J.~Yu, A.~Hu, G.~Li, and L.~Peng, ``A robust rf fingerprinting approach using
  multisampling convolutional neural network,'' \emph{IEEE Internet of Things
  Journal}, vol.~6, no.~4, pp. 6786--6799, 2019.

\bibitem{roy2019rf}
D.~Roy, T.~Mukherjee, M.~Chatterjee, and E.~Pasiliao, ``Rf transmitter
  fingerprinting exploiting spatio-temporal properties in raw signal data,'' in
  \emph{2019 18th IEEE International Conference On Machine Learning And
  Applications (ICMLA)}.\hskip 1em plus 0.5em minus 0.4em\relax IEEE, 2019, pp.
  89--96.

\bibitem{roy2019rfal}
D.~Roy, T.~Mukherjee, M.~Chatterjee, E.~Blasch, and E.~Pasiliao, ``Rfal:
  Adversarial learning for rf transmitter identification and classification,''
  \emph{IEEE Transactions on Cognitive Communications and Networking}, vol.~6,
  no.~2, pp. 783--801, 2019.

\bibitem{geffner2022introduction}
X.~Geffner and A.~Bazzan, \emph{Introduction to Semi-Supervised
  Learning}.\hskip 1em plus 0.5em minus 0.4em\relax Springer Nature, 2022.

\bibitem{chebli2018semi}
A.~Chebli, A.~Djebbar, and H.~F. Marouani, ``Semi-supervised learning for
  medical application: A survey,'' in \emph{2018 International Conference on
  Applied Smart Systems (ICASS)}.\hskip 1em plus 0.5em minus 0.4em\relax IEEE,
  2018, pp. 1--9.

\bibitem{chen2019semisupervised}
K.~Chen, L.~Yao, D.~Zhang, X.~Wang, X.~Chang, and F.~Nie, ``A semisupervised
  recurrent convolutional attention model for human activity recognition,''
  \emph{IEEE transactions on neural networks and learning systems}, vol.~31,
  no.~5, pp. 1747--1756, 2019.

\bibitem{o2017semi}
T.~J. O'Shea, N.~West, M.~Vondal, and T.~C. Clancy, ``Semi-supervised radio
  signal identification,'' in \emph{2017 19th International Conference on
  Advanced Communication Technology (ICACT)}.\hskip 1em plus 0.5em minus
  0.4em\relax IEEE, 2017, pp. 33--38.

\bibitem{li2018generative}
M.~Li, O.~Li, G.~Liu, and C.~Zhang, ``Generative adversarial networks-based
  semi-supervised automatic modulation recognition for cognitive radio
  networks,'' \emph{Sensors}, vol.~18, no.~11, p. 3913, 2018.

\bibitem{zhou2020generative}
H.~Zhou, L.~Jiao, S.~Zheng, L.~Yang, W.~Shen, and X.~Yang, ``Generative
  adversarial network-based electromagnetic signal classification: A
  semi-supervised learning framework,'' \emph{China Communications}, vol.~17,
  no.~10, pp. 157--169, 2020.

\bibitem{dong2021ssrcnn}
Y.~Dong, X.~Jiang, L.~Cheng, and Q.~Shi, ``Ssrcnn: A semi-supervised learning
  framework for signal recognition,'' \emph{IEEE Transactions on Cognitive
  Communications and Networking}, 2021.

\bibitem{shorten2019survey}
C.~Shorten and T.~M. Khoshgoftaar, ``A survey on image data augmentation for
  deep learning,'' \emph{Journal of Big Data}, vol.~6, no.~1, pp. 1--48, 2019.

\bibitem{huang2019data}
L.~Huang, W.~Pan, Y.~Zhang, L.~Qian, N.~Gao, and Y.~Wu, ``Data augmentation for
  deep learning-based radio modulation classification,'' \emph{IEEE Access},
  vol.~8, pp. 1498--1506, 2019.

\bibitem{wang2021generative}
Z.~Wang, Q.~She, and T.~E. Ward, ``Generative adversarial networks in computer
  vision: A survey and taxonomy,'' \emph{ACM Computing Surveys (CSUR)},
  vol.~54, no.~2, pp. 1--38, 2021.

\bibitem{tang2018digital}
B.~Tang, Y.~Tu, Z.~Zhang, and Y.~Lin, ``Digital signal modulation
  classification with data augmentation using generative adversarial nets in
  cognitive radio networks,'' \emph{IEEE Access}, vol.~6, pp. 15\,713--15\,722,
  2018.

\bibitem{patel2020data}
M.~Patel, X.~Wang, and S.~Mao, ``Data augmentation with conditional gan for
  automatic modulation classification,'' in \emph{Proceedings of the 2nd ACM
  Workshop on wireless security and machine learning}, 2020, pp. 31--36.

\bibitem{lee2021uniqgan}
I.~Lee and W.~Lee, ``Uniqgan: Unified generative adversarial networks for
  augmented modulation classification,'' \emph{IEEE Communications Letters},
  vol.~26, no.~2, pp. 355--358, 2021.

\bibitem{van2020survey}
J.~E. Van~Engelen and H.~H. Hoos, ``A survey on semi-supervised learning,''
  \emph{Machine Learning}, vol. 109, no.~2, pp. 373--440, 2020.

\bibitem{oliver2018realistic}
A.~Oliver, A.~Odena, C.~Raffel \emph{et~al.}, ``Realistic evaluation of deep
  semi-supervised learning algorithms,'' \emph{arXiv preprint
  arXiv:1804.09170}, 2018.

\bibitem{ouali2020overview}
Y.~Ouali, C.~Hudelot, and M.~Tami, ``An overview of deep semi-supervised
  learning,'' \emph{arXiv preprint arXiv:2006.05278}, 2020.

\bibitem{sajjadi2016regularization}
M.~Sajjadi, M.~Javanmardi, and T.~Tasdizen, ``Regularization with stochastic
  transformations and perturbations for deep semi-supervised learning,''
  \emph{Advances in neural information processing systems}, vol.~29, pp.
  1163--1171, 2016.

\bibitem{samuli2017temporal}
L.~Samuli and A.~Timo, ``Temporal ensembling for semi-supervised learning,'' in
  \emph{International Conference on Learning Representations (ICLR)}, vol.~4,
  no.~5, 2017, p.~6.

\bibitem{xie2019unsupervised}
Q.~Xie, Z.~Dai, E.~Hovy, M.-T. Luong, and Q.~V. Le, ``Unsupervised data
  augmentation for consistency training,'' \emph{arXiv preprint
  arXiv:1904.12848}, 2019.

\bibitem{berthelot2019remixmatch}
D.~Berthelot, N.~Carlini, E.~D. Cubuk, A.~Kurakin, K.~Sohn, H.~Zhang, and
  C.~Raffel, ``Remixmatch: Semi-supervised learning with distribution alignment
  and augmentation anchoring,'' \emph{arXiv preprint arXiv:1911.09785}, 2019.

\bibitem{sohn2020fixmatch}
K.~Sohn, D.~Berthelot, C.-L. Li \emph{et~al.}, ``{FixMatch: Simplifying
  semi-supervised learning with consistency and confidence},'' \emph{arXiv
  preprint arXiv:2001.07685}, 2020.

\bibitem{goldsmith2005wireless}
A.~Goldsmith, \emph{Wireless communications}.\hskip 1em plus 0.5em minus
  0.4em\relax Cambridge university press, 2005.

\bibitem{hanna2019deep}
S.~S. Hanna and D.~Cabric, ``Deep learning based transmitter identification
  using power amplifier nonlinearity,'' in \emph{2019 International Conference
  on Computing, Networking and Communications (ICNC)}.\hskip 1em plus 0.5em
  minus 0.4em\relax IEEE, 2019, pp. 674--680.

\bibitem{aghasi2007modified}
A.~R. Aghasi, H.~Karkhaneh, and A.~Ghorbani, ``A modified model and
  linearization method for solid state power amplifier,'' \emph{Analog
  Integrated Circuits and Signal Processing}, vol.~51, no.~2, pp. 81--88, 2007.

\bibitem{siddik2021wideft}
A.~B. Siddik, D.~Drake, T.~Wilkinson, P.~L. De~Leon, S.~Sandoval, and
  M.~Campos, ``{WIDEFT}: A corpus of radio frequency signals for wireless
  device fingerprint research,'' in \emph{2021 IEEE International Symposium on
  Technologies for Homeland Security (HST)}.\hskip 1em plus 0.5em minus
  0.4em\relax IEEE, 2021, pp. 1--7.

\bibitem{ioffe2015batch}
S.~Ioffe and C.~Szegedy, ``Batch normalization: Accelerating deep network
  training by reducing internal covariate shift,'' in \emph{International
  conference on machine learning}.\hskip 1em plus 0.5em minus 0.4em\relax PMLR,
  2015, pp. 448--456.

\bibitem{nair2010rectified}
V.~Nair and G.~E. Hinton, ``Rectified linear units improve restricted boltzmann
  machines,'' in \emph{Icml}, 2010.

\bibitem{simonyan2014very}
K.~Simonyan and A.~Zisserman, ``Very deep convolutional networks for
  large-scale image recognition,'' \emph{arXiv preprint arXiv:1409.1556}, 2014.

\bibitem{szegedy2015going}
C.~Szegedy, W.~Liu, Y.~Jia, P.~Sermanet, S.~Reed, D.~Anguelov, D.~Erhan,
  V.~Vanhoucke, and A.~Rabinovich, ``Going deeper with convolutions,'' in
  \emph{Proceedings of the IEEE conference on computer vision and pattern
  recognition}, 2015, pp. 1--9.

\bibitem{he2016deep}
K.~He, X.~Zhang, S.~Ren, and J.~Sun, ``Deep residual learning for image
  recognition,'' in \emph{Proceedings of the IEEE conference on computer vision
  and pattern recognition}, 2016, pp. 770--778.

\bibitem{he2019bag}
T.~He, Z.~Zhang, H.~Zhang, Z.~Zhang, J.~Xie, and M.~Li, ``Bag of tricks for
  image classification with convolutional neural networks,'' in
  \emph{Proceedings of the IEEE/CVF Conference on Computer Vision and Pattern
  Recognition}, 2019, pp. 558--567.

\bibitem{tensorflow2015}
\BIBentryALTinterwordspacing
M.~Abadi, A.~Agarwal, P.~Barham \emph{et~al.}, ``{TensorFlow}: Large-scale
  machine learning on heterogeneous systems,'' 2015, software available from
  tensorflow.org. [Online]. Available: \url{https://www.tensorflow.org/}
\BIBentrySTDinterwordspacing

\bibitem{kingma2014adam}
D.~P. Kingma and J.~Ba, ``Adam: A method for stochastic optimization,''
  \emph{arXiv preprint arXiv:1412.6980}, 2014.

\end{thebibliography}

\end{document}